\documentclass[%
 reprint,
 amsmath,amssymb,
 aps,superscriptaddress
]{revtex4-2}
\usepackage{graphicx}
\usepackage{dcolumn}
\usepackage{bm}
\usepackage[normalem]{ulem}
\usepackage{xcolor}
\usepackage{mathrsfs}
\usepackage{inputenc}
\usepackage[version=4]{mhchem}
\usepackage[slantedGreek]{mathpazo}
\usepackage{hyperref}
\usepackage{float}
\usepackage{soul}
\usepackage{comment}
\begin{document}
\newcommand{\rb}[1]{\textcolor{red}{\it#1}}
\newcommand{\rbout}[1]{\textcolor{red}{\sout{#1}}}

\preprint{APS/123-QED}

\title{Accurate \textit{ab initio} calculations of the ground state properties of singly charged alkali metal hydrides}

\author{Ankush Thakur}
\email{ankush\_t@ph.iitr.ac.in}
\affiliation{Department of Physics, Indian Institute of Technology Roorkee, Roorkee-247667, India}
\author{Renu Bala}
\email{renub@umk.pl}
\affiliation{Institute of Physics, Faculty of Physics, Astronomy and Informatics, Nicolaus Copernicus University, Grudziadzka 5, 87-100 Toru\'n, Poland}
\author{H. S. Nataraj}
\affiliation{Department of Physics, Indian Institute of Technology Roorkee, Roorkee-247667, India}

\begin{abstract}

In this article, we investigate the ground state electronic properties of singly charged alkali metal hydrides using the single-reference coupled-cluster method with single and double excitations (CCSD), together with perturbative triple excitations (CCSD(T)). A large relativistic core-valence quintuple-zeta basis set has been employed. The calculated spectroscopic constants and molecular properties are compared with existing data in the literature. Further, the relevant vibrational parameters are determined by solving the vibrational Schr\"odinger equation utilizing potential energy curves and permanent dipole moment curves. Thereafter, spontaneous and black-body radiation induced transition rates are calculated to obtain the lifetimes of the vibrational states.

\begin{description}
\item[Keywords]
alkali metal hydrides, potential energy curves, spectroscopic constants, molecular properties, vibrational parameters, transition rates, lifetimes.
\end{description}

\end{abstract}

\maketitle

\section{\label{sec:section1}Introduction }

Alkali metal hydrides, including both ionic and neutral species, are of significant interest in astrochemistry owing to their potential role in the formation and chemical evolution of the molecular universe~\cite{Levine_1985}. In particular, alkali metal atoms readily react with molecular hydrogen to form alkali metal hydrides, which are relevant to the processes involved in star formation~\cite{Lin_1989,Hack_1999}. The spectral signatures observed in brown dwarfs and extrasolar giant planets predominantly arise from resonance lines of alkali metal species interacting with helium and molecular hydrogen rich atmospheres~\cite{Sharp_2007}. Furthermore, the ionization balance of planetary atmospheres, dwarf stars, and the interstellar medium is affected by collisions between alkali metal atoms and protons, as well as by processes involving their molecular ions~\cite{Langer_1990,Watson_1996,Burrows_2001}. Collisions between alkali metal atoms and hydrogen atoms are also of interest in ultracold physics, where proposals have been made to realize Bose-Einstein condensation of atomic hydrogen through enhanced cooling using a buffer gas of alkali metal atoms~\cite{Cote_2000,Derevianko_2001}.\\

The alkali metal hydride cations (AlkH$^+$, Alk\,=\,Li, Na, K, Rb, Cs, and Fr) can be produced in the laboratory, typically through ion-molecule reactions or plasma sources, followed by isolation in sufficient quantities for precise measurements~\cite{Krohn_2001,Zajfman_2003}. To our knowledge, the only available experimental data for AlkH$^+$ pertain to an excited state of CsH$^+$ and it has been extracted from differential cross-section measurements~\cite{Scheidt_1978}. The scarcity of experimental data on AlkH$^+$ molecular ions is due to the inherent challenges involved in their preparation and characterization.\\

On the theoretical front, a few studies of AlkH$^+$ ions have been reported in the literature. For example, the model potential-type method~\cite{Magnier_2004,Magnier_2005,Magnier_2006}, various pseudopotential approaches~\cite{Aymar_2009,Korek_2008,Laszlo_1982,Ghanmi_2025}, and the pseudonatural orbital configuration interaction method~\cite{Rosmus_1977} have been employed to compute their spectroscopic parameters. An \textit{ab initio} multireference single- and double-excitation configuration interaction (MRD-CI) study of the ground and excited electronic states of the AlkH$^+$ series, excluding FrH$^+$, has been reported in Ref.~\cite{Yan_2012}. Recently, Wang \textit{et al.}~\cite{Wang_2025} have studied the photodissociation cross sections of NaH$^+$ to CsH$^+$ using potential energy curves (PECs) and transition dipole moments (TDMs) calculated with the MRD-CI method. To provide accurate data for interpreting the results of both laboratory and astrophysical investigations, we have employed the fully-relativistic coupled-cluster method restricted to single, double, and perturbative triple excitations, which accounts for extensive electron correlation effects in conjunction with the largest available relativistic quintuple-zeta basis set in the DIRAC suite~\cite{DIRAC}.\\

The present work reports the ground state spectroscopic parameters, namely the equilibrium bond length ($R_e$), dissociation energy ($D_e$), harmonic and anharmonic frequencies ($\omega_e$ and $\omega_e x_e$), and rotational constants ($B_e$ and $\alpha_e$), together with the valence molecular properties, including the permanent electric dipole moments (PDMs), the components of the static electric dipole polarizability ($\alpha_\parallel$ and $\alpha_\perp$), and the quadrupole moments (QMs) of AlkH$^+$ molecular ions. The variation of these properties as a function of internuclear distance has also been examined. Furthermore, we have carried out calculations of vibrational parameters such as vibrational wavefunctions, energies ($E_v$), TDMs between vibrational states, vibrationally coupled rotational constants ($B_v$), spontaneous and black-body radiation (BBR) induced transition rates, and lifetimes of the vibrational states of the aforementioned ions. \\

The paper is structured as follows: After the introduction, Section~\ref{sec:section2} gives the theoretical and calculation details, which is followed by Section~\ref{sec:section3} that reports the detailed discussion of the computed results. Finally, the work is summarized in Section~\ref{sec:section4}.

\section{\label{sec:section2}Theory and calculational details}

\textit{Ab initio} electronic structure calculations have been performed employing a four-component relativistic formalism at the Dirac-Hartree-Fock (DHF), coupled-cluster singles and doubles (CCSD), and perturbative triples (CCSD(T)) levels of theory using the DIRAC23~\cite{DIRAC} software package. The largest available relativistic Dyall core-valence quintuple-zeta basis set (dyall.cv5z)~\cite{Reitsma_2025} for the s-block elements in the DIRAC database has been utilized for these calculations, with the details of the basis functions given in Table~\ref{table-I}. To make the calculations computationally feasible, only molecular orbitals with energies ranging from \(-5\) $E_h$ to \(13\) $E_h$ have been included in the correlation treatment. Despite this restriction, the active correlation space is still quite large, as indicated in Table~\ref{table-II}. Potential energy curves (PECs) are computed over the range of \(1\) to \(30\) {\AA}, with a finer step size of \(0.001\) {\AA} used near the equilibrium region. The diatomic constants are then derived from the PECs using the OpenMolcas package~\cite{OPENMOLCAS}. The finite-field method~\cite{Maroulis_1998,Thakur_2024} has been employed to determine the molecular response properties by evaluating the derivatives of the energy with respect to a perturbative electric field. The strength of the electric field considered as a perturbation is taken in the range of \(-1\) $\times$ $10^{-4}$ to \(1\) $\times$ $10^{-4}$ au. The origin of the coordinate system is taken to be at the center-of-mass of the molecular ions.

\begin{table}[ht]
    \centering
    \caption{\label{table-I}Details of the dyall.cv5z basis functions~\cite{Reitsma_2025}.}
    \begin{tabular}{c c c c}
    \hline\hline
       Atom &&& Basis functions \\
       \hline
       H &&& 13s, 4p, 3d, 2f, 1g \\
       Li &&& 24s, 14p, 7d, 5f, 3g \\
       Na &&& 28s, 18p, 11d, 8f, 5g, 2h \\
       K &&& 35s, 26p, 11d, 9f, 6g, 3h \\
       Rb &&& 40s, 33p, 21d, 10f, 7g, 4h, 1i \\
      Cs &&& 43s, 38p, 26d, 10f, 7g, 4h, 1i \\
      Fr &&& 43s, 43p, 28d, 19f, 7g, 4h, 1i \\
       \hline\hline
    \end{tabular}
    \label{tab:my_label}
\end{table}

\begin{table}[ht]
    \centering
    \caption{\label{table-II}Number of active electrons and virtual orbitals within the energy range of \(-5\) $E_h$ to \(13\) $E_h$.}
    \begin{tabular}{c c c c c c c}
    \hline\hline
    Molecule &&& Active electrons && Virtual orbitals \\
    \hline
    LiH$^+$ &&& 3 && 155 \\
    NaH$^+$ &&& 9 && 171 \\
    KH$^+$ &&& 9 && 223 \\
    RbH$^+$ &&& 19 && 234 \\
    CsH$^+$ &&& 19 && 256 \\
    FrH$^+$ &&& 19 && 254 \\
   \hline\hline
    \end{tabular}
    \label{tab:my_label}
\end{table}

With the internuclear axis of the molecular systems defined along the z-axis, we have determined the parallel (\( \alpha_\parallel \equiv \alpha_{zz}\)) and perpendicular components (\( \alpha_\perp \equiv \alpha_{xx} = \alpha_{yy}\)) of static electric dipole polarizability. At large internuclear distances, they approach their asymptotic limits as given in terms of their atomic polarizabilities as~\cite{Silberstein_1917,Jensen_2002}:
\begin{equation}
    \alpha_{\parallel}(R) \approx \alpha_{Alk^+} + \alpha_H + \frac{4 \, \alpha_{Alk^+} \, \alpha_H}{R^3} + \frac{4 \, (\alpha_{Alk^+} + \alpha_H)\alpha_{Alk^+}\, \alpha_H}{R^6},
    \label{eq:1}
\end{equation}
\begin{equation}
    \alpha_{\perp}(R) \approx \alpha_{Alk^+} + \alpha_H - \frac{2 \, \alpha_{Alk^+} \, \alpha_H}{R^3} + \frac{(\alpha_{Alk^+} + \alpha_H)\alpha_{Alk^+}\, \alpha_H}{R^6}.
    \label{eq:2}
\end{equation}

The average ($\bar{\alpha}$) and anisotropic ($\gamma$) components of the electric dipole polarizability are given by,
\begin{equation}
    \bar{\alpha}=(\alpha_\parallel + 2\alpha_\perp)/3,
\end{equation}
\begin{equation}
    \gamma=\alpha_\parallel - \alpha_\perp.
\end{equation}

The traceless quadrupole moment (QM) tensor, $\Theta_{mn}$, can be expressed as~\cite{Buckingham_1959},
\begin{eqnarray}
\Theta_{mn}=\frac{1}{2}\,\sum_{i}e_i(3r_{i_m}\,r_{i_n}\,-\,r_{i}^2\,\delta_{mn}),
\end{eqnarray}
where $m$ and $n$ denote the Cartesian components, and the summation index $i$ runs over all charged particles in the system. The $zz$-component of the traceless QM tensor, \emph{viz}. $\Theta_{zz}$, is related to the other diagonal components through the relation, 
\begin{eqnarray}
\Theta_{zz}\,=-(\,\Theta_{xx}\,+\,\Theta_{yy}).
\end{eqnarray}
For linear molecules, $\Theta_{xx}$ = $\Theta_{yy}$.\\

Further, the vibrational Schr\"odinger equation has been solved using the PECs shown in Fig.~\ref{fig:FIG1} and PDM curves shown in Fig.~\ref{fig:FIG2}. The resulting vibrational wavefunctions, vibrational energy levels ($E_v$), vibrationally coupled rotational constants ($B_v$), and transition dipole moments (TDMs) between different vibrational states have been obtained using Le Roy’s LEVEL program~\cite{LEVEL}. The spontaneous and BBR-induced transition rates at room temperature ($\mathrm{T}$ = \(300\) $\mathrm{K}$) have been calculated using the equations~\cite{Kotochigova_2005,Thakur_2026},
\begin{eqnarray}{}  
\Gamma_{v, J}^{spon}\,=\,\sum\limits_{v^{''}, J^{''}}\Gamma^{emis}(v, J\,\rightarrow\,v^{''}, J^{''}) 
 \end{eqnarray}
and
 \begin{eqnarray}{} 
 \Gamma_{v, J}^{BBR}\,&=& \,\sum\limits_{v^{''}, J^{''}}\bar{n}(\omega)\,\Gamma^{emis}(v, J\,\rightarrow\,v^{''}, J^{''})\nonumber\\
 &+&\sum\limits_{v^{'}, J^{'}}\bar{n}(\omega)\,\Gamma^{abs}(v, J\,\rightarrow\,v^{'}, J^{'}),
  \end{eqnarray}
respectively. Here, ($v^{''}, J^{''}$) and ($v^{'}, J^{'}$) represent the ro-vibrational states with energies lower and higher, respectively, than that of the ($v, J$) state. The average number of photons at frequency $\omega$, denoted by $\bar{n}(\omega)$, is given by,
\begin{eqnarray}{}
\bar{n}(\omega)\,=\,\frac{1}{e^{(\hbar\omega/k_{B}T)}-1}\,,
\end{eqnarray}
where $\hbar\omega\,=\,\arrowvert E_{v, J}-E_{\widetilde{v}, \widetilde{J}}\arrowvert$ denotes the energy separation between the two ro-vibrational levels involved, and $k_B$ is the Boltzmann constant. For emission, ($\widetilde v, \widetilde J$) corresponds to the lower energy ro-vibrational level ($v^{''}, J^{''}$), while for absorption it represents the higher energy level ($v^{'}, J^{'}$).\\

The emission and absorption rates appearing in the above expressions for $\Gamma^{spon}$ and $\Gamma^{BBR}$ are calculated using the vibrational energy differences and TDM values as,
\begin{eqnarray}{}
\Gamma^{emis/abs} [(v, J)\,\rightarrow\,(\widetilde v, \widetilde J)]\,=\nonumber\\ \,\frac{8\pi}{3\epsilon_0}\frac{\omega^3}{h c^3}\, [TDM_{(v, J)\rightarrow (\widetilde v, \widetilde J)}]^2.
\end{eqnarray}

Finally, the total lifetime ($\tau$) of the ro-vibrational state is obtained as,
\begin{eqnarray}{}
\tau\,=\,\frac{1}{\Gamma^{spon}\,+\,\Gamma^{BBR}}.
\end{eqnarray}

\section{\label{sec:section3}Results and discussion}

\subsection{Diatomic constants}

\begin{figure}[htbp]
    \includegraphics[width=1.03\linewidth]{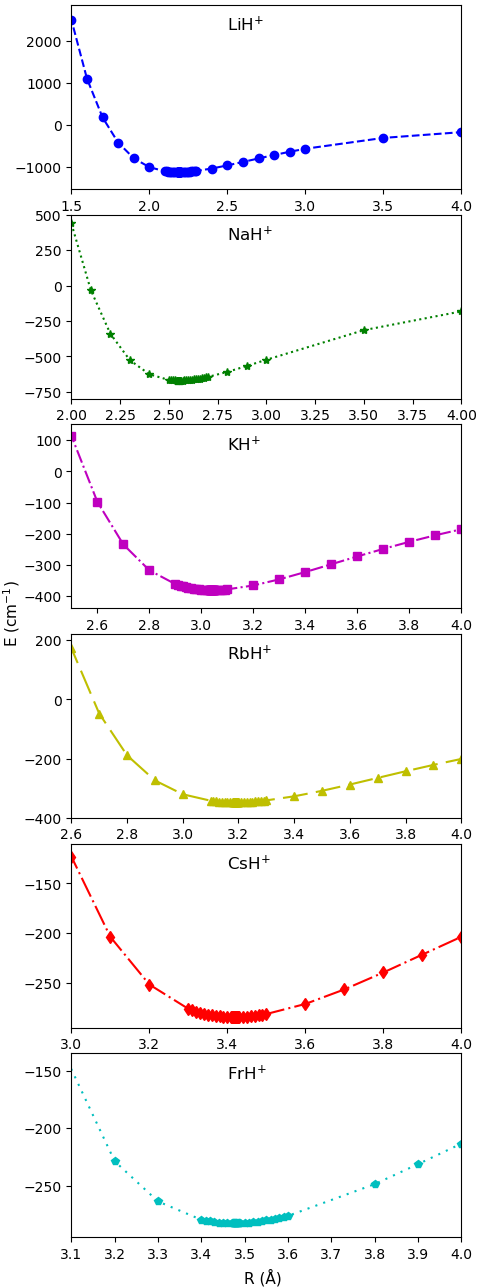}
    \caption{\label{fig:FIG1}PECs for AlkH$^+$ molecular ions computed at the CCSD(T) level of theory using cv5z basis set, relative to the energies at the dissociation limit.}
\end{figure}

\begin{table*}[htbp]
\begin{ruledtabular}
\begin{center}
\caption{\label{table-III}
Spectroscopic constants for the ground electronic states of AlkH$^+$ molecular ions computed in this work using cv5z basis set, along with available literature values. The results presented in bold fonts are our recommended values.}
\begin{tabular}{ccccccc}
 Method & $R_e$ & $D_e$ & $B_e$  & $\alpha_e$  & $\omega_e$ & $\omega_e x_e$ \\
&  (\AA) & (eV) & (cm$^{-1}$) & (cm$^{-1}$) & (cm$^{-1}$) & (cm$^{-1}$) \\
\hline
\multicolumn{7}{c}{\textbf{LiH$^+$}}\\
\hline

DHF & 2.214 & 0.1319 & 3.898 & 0.4787 & 428.80 & 43.818 \\
CCSD & 2.186 & 0.1402 & 3.995 & 0.4793 & 445.59 & 45.606 \\
\textbf{CCSD(T)} & \textbf{2.185} & \textbf{0.1404} & \textbf{3.997} & \textbf{0.4793} & \textbf{445.87} & \textbf{45.614} \\
MRD-CI~\cite{Yan_2012} & 2.16 & 0.151 & - & - & 612.99 & - \\
~\cite{Aymar_2009}$\footnote{Pseudopotential approach}$ & 2.17 & 0.1426 & - & - & - & - \\
RCCSD(T)~\cite{Cheng_2007}$\footnote{RCCSD(T): Restricted CCSD(T)}$ & 2.261 & - & 3.7416 & - & - & - \\
~\cite{Magnier_2004}$\footnote{Model potential-type method}$ & 2.19 & 0.130 & - & - & 417.03 & - \\
PNO-CI~\cite{Rosmus_1977} & 2.191 & 0.13 & 3.98 & 0.484 & 443.2 & 42.1 \\
 \hline
 \multicolumn{7}{c}{\textbf{NaH$^+$}}\\
\hline
DHF & 2.661 & 0.0673 & 2.486 & 0.4425 & 279.08 & 39.382  \\
CCSD & 2.555 & 0.0826 & 2.654 & 0.3826 & 318.31 & 46.193 \\
\textbf{CCSD(T)} & \textbf{2.552} & \textbf{0.0833} & \textbf{2.663} & \textbf{0.3834} & \textbf{320.18} & \textbf{46.466} \\
MRD-CI~\cite{Yan_2012} & 2.65 & 0.125 & - & - & 451.67 & -\\
~\cite{Aymar_2009}$\footnotemark[1]$ & 2.57 & 0.083 & - & - & - & -\\
RCCSD(T)~\cite{Cheng_2007} & 2.698 & - & 2.398 & - & - & - \\
~\cite{Valance_1978}$\footnotemark[1]$ & 3.07 & 0.020 & - & - & - & - \\
~\cite{Watanabe_2002}$\footnotemark[1]$ & 2.46 & 0.1034 & - & - & - & - \\
~\cite{Magnier_2005}$\footnotemark[3]$ & 2.59 & 0.062 & - & - & 266.70 & -\\
PNO-CI~\cite{Rosmus_1977} & 2.593 & 0.12 & 2.59 & 0.395 & 302.4 & 40.6 \\
\hline
 \multicolumn{7}{c}{\textbf{KH$^+$}}\\
\hline
DHF & 3.247 & 0.0316 & 1.453 & 0.4042 & 157.35 & 30.345 \\
CCSD & 3.055 & 0.0460 & 1.864 & 0.3785 & 222.03 & 37.137 \\
\textbf{CCSD(T)} & \textbf{3.044} & \textbf{0.0471} & \textbf{1.878} & \textbf{0.3768} & \textbf{225.35} & \textbf{37.242} \\
MRD-CI~\cite{Yan_2012} & 3.18 & 0.056 & - & - & 201.64 & - \\
~\cite{Aymar_2009}$\footnotemark[1]$ & 3.00 & 0.062 & - & - & - & -  \\
~\cite{Valance_1978}$\footnotemark[1]$ & 5.29 & 0.025 & - & - & - & -  \\
~\cite{Watanabe_2002}$\footnotemark[1]$ & 2.65 & 0.1361 & - & - & - & -  \\
~\cite{Magnier_2006}$\footnotemark[3]$ & 2.92 & 0.048 & - & - & 371.80 & -  \\
\hline
 \multicolumn{7}{c}{\textbf{RbH$^+$}}\\
\hline
DHF & 3.484 & 0.0241 & 1.230 & 0.3621 & 130.47 & 25.929 \\
CCSD & 3.204 & 0.0417 & 1.672 & 0.3354 & 198.38 & 31.605 \\
\textbf{CCSD(T)} & \textbf{3.189} & \textbf{0.0430} & \textbf{1.687} & \textbf{0.3320} & \textbf{202.07} & \textbf{31.682} \\
MRD-CI~\cite{Yan_2012}& 3.23 & 0.052 & - & - & 241.97 & -  \\
~\cite{Aymar_2009}$\footnotemark[1]$ & 3.08 & 0.6824 & - & - & - & -  \\
~\cite{Valance_1978}$\footnotemark[1]$ & 7.09 & 0.012 & - & - & - & -  \\
~\cite{Korek_2008}$\footnotemark[1]$ & 3.09 & - & 1.774 & - & 238.6 & -  \\
~\cite{Laszlo_1982}$\footnotemark[1]$ & 3.09 & 0.043 & 1.76 & - & 223 $\pm$ 10 & -  \\
\hline
\multicolumn{7}{c}{\textbf{CsH$^+$}}\\
\hline
DHF & 3.789 & 0.0175 & 1.021 & 0.3211 & 104.49 & 19.550 \\
CCSD &  3.439 & 0.0337 & 1.296 & 0.2757 & 161.98 & 27.101 \\
\textbf{CCSD(T)} & \textbf{3.418} & \textbf{0.0353} & \textbf{1.316} & \textbf{0.2731} & \textbf{166.50} & \textbf{27.315} \\
MRD-CI~\cite{Yan_2012} & 3.49 & 0.067 & - & - & 250.03 & - \\
~\cite{Aymar_2009}$\footnotemark[1]$ & 3.16 & 0.0809 & - & - & - & -  \\
~\cite{Valance_1978}$\footnotemark[1]$ & 9.31 & 0.030 & - & - & - & -  \\
~\cite{Laszlo_1982}$\footnotemark[1]$ & 3.25 & 0.035 & 1.59 & - & 204 $\pm$ 10 & - \\
\hline
\multicolumn{7}{c}{\textbf{FrH$^+$}}\\
\hline
DHF & 3.897 & 0.0156 & 0.957 & 0.1790 & 96.59 & 6.825 \\
CCSD & 3.498 & 0.0334 & 1.249 & 0.1162 & 159.62 & 14.110   \\
\textbf{CCSD(T)} & \textbf{3.476} & \textbf{0.0351} & \textbf{1.270} & \textbf{0.1123} & \textbf{164.42} & \textbf{15.918}  \\
~\cite{Ghanmi_2025}$\footnotemark[1]$ & 7.334 & 0.00124 & 0.309 & - & 5.61 & 0.15 \\
\end{tabular}
\end{center}
\end{ruledtabular}
\end{table*}

The computed PECs for all the AlkH$^+$ molecular ions at the CCSD(T) level of theory are presented in Fig.~\ref{fig:FIG1}. Full PECs in numerical form are collected in Supplementary Table S1. The corresponding spectroscopic constants such as $R_e$, $D_e$, $\omega_e$, $\omega_e x_e$, $B_e$, and $\alpha_e$ are summarized in Table~\ref{table-III}, along with the results reported in the literature. The values computed at the CCSD(T) level shown in bold font in the table represent our recommended results.\\

As shown in Table~\ref{table-III}, the equilibrium internuclear distances of the investigated molecular ions increase systematically with the increasing mass of the alkali metal atom. This leads to a decrease in $B_e$ and $\omega_e$ from lighter to heavier molecular systems. Furthermore, the potential well depth decreases from \(0.1404\) eV for LiH$^+$ to \(0.0351\) eV for FrH$^+$. Compared to the DHF results, the electron correlation contribution to the computed $D_e$ at the CCSD(T) level is observed to increase across the AlkH$^+$ series, reaching a maximum contribution of \(125\)\% for FrH$^+$. The inclusion of correlation effects increases $\omega_e x_e$ while lowering $R_e$ for all molecular ions. It also decreases $\alpha_e$, except in the case of LiH$^+$ ion.\\

A comparison of $R_e$, $D_e$, and $\omega_e$ values for LiH$^+$ to CsH$^+$ molecular ions obtained at the CCSD(T) level with previously reported multireference single- and double-excitation configuration interaction (MRD-CI)~\cite{Yan_2012} results yields absolute differences ranging from \(0.025\)-\(0.136\) {\AA} for $R_e$, \(0.009\)-\(0.0417\) eV for $D_e$, and \(23.71\)-\(167.12\) cm$^{-1}$ for $\omega_e$. Further, the differences between our recommended results computed at the CCSD(T) level and those obtained using the pseudonatural orbital configuration interaction (PNO-CI) method~\cite{Rosmus_1977} remain within \(6\)\% for all constants of LiH$^+$ and NaH$^+$, except for $D_e$ and $\omega_e x_e$. The observed differences may be attributed to distinct choices of correlation spaces, basis sets, and computational methodologies.\\

Most existing theoretical studies have employed model potential-type methods~\cite{Magnier_2004,Magnier_2005,Magnier_2006} and various pseudopotential approaches~\cite{Korek_2008,Laszlo_1982,Aymar_2009,Ghanmi_2025}, which have been limited to the quasi-one-electron approximation. A comprehensive treatment of electron correlation effects requires methods that go beyond this approximation. The present \textit{ab initio} many-electron calculations, performed using a highly correlated, fully relativistic coupled-cluster method and a larger basis set, therefore provide reliable results for the diatomic constants of AlkH$^+$ molecular ions. Consequently, we refrain from explicitly comparing our recommended values with those obtained using the aforementioned methods. However, they are presented in the Table~\ref{table-III}.\\

To examine the dissociation of the AlkH$^+$ molecular ions into Alk$^+$ and H atoms, we have compared the molecular energies at the dissociation limit ($R$ = \(30\) {\AA}) with the sum of the atomic energies of Alk$^+$ and H atoms and found that the two energy values differ by only \(0.1\) micro Hartree, thereby confirming our presumption.

\subsection{Molecular properties}

\begin{figure}[h]
    \includegraphics[width=1.03\linewidth]{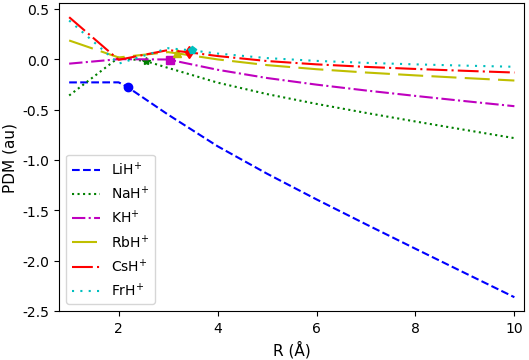}
    \caption{\label{fig:FIG2}PDM curves for AlkH$^+$ molecular ions computed at the CCSD(T) level of theory using cv5z basis set. The markers on these curves indicate PDM values at $R_e$.}
\end{figure}

\begin{figure}[h]
    \includegraphics[width=1.03\linewidth]{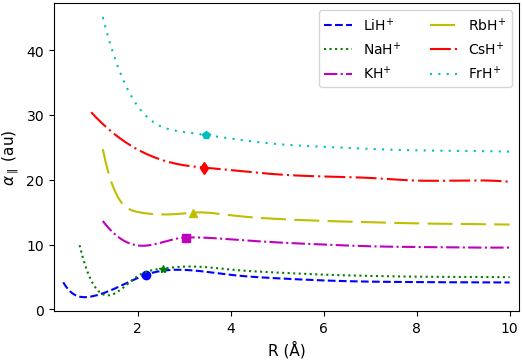}
    \caption{\label{fig:FIG3}The $\alpha_\parallel$ for AlkH$^+$ molecular ions computed at the CCSD(T) level of theory using cv5z basis set. The markers on these curves indicate $\alpha_\parallel$ values at $R_e$.}
\end{figure}

\begin{figure}[]
    \includegraphics[width=1.03\linewidth]{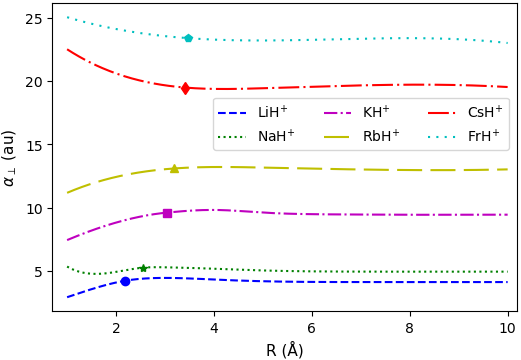}
    \caption{\label{fig:FIG4}The $\alpha_\perp$ for AlkH$^+$ molecular ions computed at the CCSD(T) level of theory using cv5z basis set. The markers on these curves indicate $\alpha_\perp$ values at $R_e$.}
\end{figure}

\begin{figure}[]
    \includegraphics[width=0.91\linewidth]{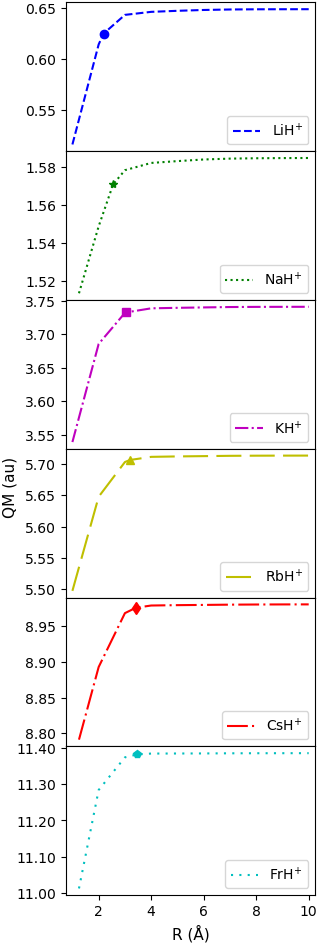}
    \caption{\label{fig:FIG5}QM curves for AlkH$^+$ molecular ions computed at the CCSD(T) level of theory using cv5z basis set. The markers on these curves indicate QM values at $R_e$.}
\end{figure}

\begin{table*}[htbp]
\caption{\label{table-IV}
Magnitude of molecular properties: electric dipole moments ($\mu_e$) and electric quadrupole moments ($\Theta_{zz}$), and components of dipole polarizabilities ($\alpha_\parallel, \alpha_\perp$, $\bar{\alpha}$ and $\gamma$); for the ground state of AlkH$^+$ molecular ions. All the properties are given in atomic units (au). The results presented in bold fonts are our recommended values.}
\begin{ruledtabular}
\begin{tabular}{cccccccc}
Molecule & Method & $\mu_e$ & $\Theta_{zz}$ & $\alpha_\parallel$ & $\alpha_\perp$ & $\bar{\alpha}$ & $\gamma$\\
\hline
\textbf{LiH$^+$} & DHF & 0.289 & 0.6248 & 5.381 & 4.199 & 4.593 & 1.182 \\
        & CCSD & 0.278 & 0.6246 & 5.363 & 4.189 & 4.580 & 1.174 \\
        & \textbf{CCSD(T)} & \textbf{0.277} & \textbf{0.6246} & \textbf{5.361} & \textbf{4.220} & \textbf{4.601} & \textbf{1.141} \\
        & \cite{Magnier_2004}$\footnote{The parallel and perpendicular components of the dipole polarizability have been extracted from the graphs given in Refs.~\cite{Magnier_2004,Magnier_2005}}^,\footnote{Model potential-type method}$ & 0.276 & - & 5.034 & 3.839 & - & - \\
        & RCCSD(T)~\cite{Cheng_2007}$\footnote{RCCSD(T): Restricted CCSD(T)}$ & 0.3049 & - & - & - & - & - \\
        \hline
\textbf{NaH$^+$} & DHF & 0.038 & 1.5550 & 6.216 & 5.060 & 5.445 & 1.156 \\
        & CCSD & 0.019 & 1.5701 & 6.287 & 5.091 & 5.490 & 1.196 \\
        & \textbf{CCSD(T)} & \textbf{0.018} & \textbf{1.5710} & \textbf{6.296} & \textbf{5.255} & \textbf{5.602} & \textbf{1.041} \\
        & \cite{Magnier_2005}$\footnotemark[1]^,\footnotemark[2]$ & - & - & 5.116 & 3.877 & - & - \\
        & RCCSD(T)~\cite{Cheng_2007}$\footnotemark[3]$ & 0.0485 & - & - & - & - & - \\
        \hline
\textbf{KH$^+$} & DHF & 0.032 & 3.7390 & 10.878 & 9.488 & 9.951 & 1.390 \\
        & CCSD & 0.007 & 3.7314 & 11.078 & 9.497 & 10.024 & 1.581 \\
        & \textbf{CCSD(T)} & \textbf{0.006} & \textbf{3.7326} & \textbf{11.085} & \textbf{9.610} & \textbf{10.102} & \textbf{1.475} \\
        \hline
\textbf{RbH$^+$} & DHF & 0.033 & 5.7333 & 14.654 & 13.029 & 13.571 & 1.625 \\
        & CCSD & 0.058 & 5.7048 & 14.825 & 12.943 & 13.570 & 1.882 \\
        & \textbf{CCSD(T)} & \textbf{0.059} & \textbf{5.7065} & \textbf{14.941} & \textbf{13.099} & \textbf{13.713} & \textbf{1.842} \\
        & \cite{Laszlo_1982}$\footnote{In Ref.~\cite{Laszlo_1982}, the PDM values were calculated with respect to the bond center. We have translated these values to the center-of-mass of the molecular ions for comparison with our results}^,\footnote{Pseudopotential approach}$ & 0.042 & - & - & - & - & - \\
        \hline
\textbf{CsH$^+$} & DHF & 0.042 & 9.0325 & 21.540 & 19.675 & 20.296 & 1.864 \\
        & CCSD & 0.068 & 8.9744 & 21.843 & 19.495 & 20.278 & 2.348 \\
        & \textbf{CCSD(T)} & \textbf{0.070} & \textbf{8.9758} & \textbf{21.894} & \textbf{19.493} & \textbf{20.293} & \textbf{2.401} \\
        & \cite{Laszlo_1982}$\footnotemark[4]^,\footnotemark[5]$ & 0.064 & - & - & - & - & - \\
        \hline
        \textbf{FrH$^+$} & DHF & 0.060 & 11.4342 & 26.290 & 24.120 & 24.843 & 2.170 \\
        & CCSD & 0.087 & 11.3793 & 26.780 & 23.612 & 24.668 & 3.168 \\
        & \textbf{CCSD(T)} & \textbf{0.089} & \textbf{11.3826} & \textbf{26.910} & \textbf{23.400} & \textbf{24.570} & \textbf{3.510} \\
\end{tabular}
\end{ruledtabular}
\end{table*}

\begin{table*}
\caption{\label{table-V}
Comparison of molecular polarizabilities of AlkH$^+$ ions at the dissociation limit with the sum of atomic polarizabilities, both calculated at the CCSD(T) level of theory. All values are given in atomic units (au).}
\begin{ruledtabular}
\begin{tabular}{cccccc}
&& \multicolumn{2}{c}{$\alpha_\parallel$} & \multicolumn{2}{c}{$\alpha_\perp$} \\
\cline{3-4}\cline{5-6}
\multicolumn{4}{c}{\vspace{-3mm}}\\
Molecule & $\alpha_{\text{Alk}^{+}}$ + $\alpha_\text{H}$ & calculated from Eq.~(\ref{eq:1}) & at ${R=30\,\text{\AA}}$ & calculated from Eq.~(\ref{eq:2}) & at ${R=30\,\text{\AA}}$ \\
\multicolumn{4}{c}{\vspace{-3mm}}\\
\hline
\multicolumn{4}{c}{\vspace{-3mm}}\\
LiH$^+$ & 4.117 & 4.117 & 4.117 & 4.117 & 4.118 \\
NaH$^+$ & 4.988 & 4.989 & 4.989 & 4.988 & 4.988 \\ 
KH$^+$ & 9.447 & 9.450 & 9.454 & 9.445 & 9.446 \\
RbH$^+$ & 13.066 & 13.071 & 12.986 & 13.063 & 13.027 \\
CsH$^+$ & 19.663 & 19.672 & 19.707 & 19.658 & 19.683 \\
FrH$^+$ & 24.291 & 24.303 & 24.320 & 24.285 & 24.297 \\
\end{tabular}
\end{ruledtabular}
\end{table*}

\begin{table*}
\caption{\label{table-VI}
Vibrational energies ($E_v$), vibrationally coupled rotational constants ($B_v$), spontaneous and BBR-induced transition rates, and lifetimes for the vibrational levels of the ground electronic states of AlkH$^+$ molecular ions.}
\begin{ruledtabular}
\begin{tabular}{ccccccc}
Molecule & $v$ & $E_v$ (cm$^{-1}$) & $B_v$ (cm$^{-1}$) & $\Gamma_{v}^{spon}$ (s$^{-1}$) & $\Gamma_{v}^{BBR}$ (s$^{-1}$) & $\tau$ (s) \\
\hline
\multicolumn{4}{c}{\vspace{-3mm}}\\
\textbf{LiH$^+$} & 0 & 211.513 & 3.748 &  $-$& 0.087 & 11.494 \\
        & 1 & 565.736 & 3.200 & 0.371 & 0.252 & 1.605 \\
        & 2 & 827.559 & 2.583 & 0.421 & 0.317 & 1.355 \\
        & 3 & 997.964 & 1.901 & 0.304 & 0.218 & 1.916 \\
        & 4 & 1088.966 & 1.186 & 0.149 & 0.099 & 4.032 \\
\hline
\multicolumn{4}{c}{\vspace{-3mm}}\\
\textbf{NaH$^+$} & 0 & 148.669 & 2.464 & $-$& 0.017 & 57.736 \\
        & 1 & 381.901 & 2.028 & 0.032 & 0.040 & 13.935 \\
        & 2 & 539.631 & 1.538 & 0.031 & 0.032 & 15.958 \\
        & 3 & 625.864 & 1.014 & 0.017 & 0.015 & 31.227 \\
\hline
\multicolumn{4}{c}{\vspace{-3mm}}\\
\textbf{KH$^+$} & 0 & 103.365 & 1.689 & $-$& 0.006 & 161.684 \\
        & 1 & 254.301 & 1.312 & 0.006 & 0.009 & 66.150 \\
        & 2 & 337.505 & 0.890 & 0.004 & 0.005 & 114.372 \\
\hline
\multicolumn{4}{c}{\vspace{-3mm}}\\
\textbf{RbH$^+$} & 0 & 93.120 & 1.520 &$-$ & 0.003 & 344.903 \\
        & 1 & 231.901 & 1.187 & 0.002 & 0.004 & 165.522 \\
        & 2 & 308.158 & 0.802 & 0.001 & 0.002 & 331.231 \\
\hline
\multicolumn{4}{c}{\vspace{-3mm}}\\
\textbf{CsH$^+$} & 0 & 83.049 & 1.317 & $-$& 0.0020 & 509.010 \\
        & 1 & 200.732 & 1.008 & 0.0012 & 0.0023 & 282.242 \\
        & 2 & 260.203 & 0.647 & 0.0007 & 0.0011 & 555.877 \\
        \hline
\textbf{FrH$^+$} & 0 & 82.157 & 1.270 & $-$& 0.0017 & 582.406 \\
         & 1 & 198.674 & 0.978 & 0.0010 & 0.0020 & 335.335 \\
         & 2 & 258.209 & 0.630 & 0.0006 & 0.0009 & 655.984 \\
\end{tabular}
\end{ruledtabular}
\end{table*}

The PDM curves for the AlkH$^+$ molecular ions computed at the CCSD(T) level are shown in Fig.~\ref{fig:FIG2} as a function of internuclear distance ($R$). These curves reveal that the PDM values for all the molecular ions decrease with increasing $R$ beyond the equilibrium bond length ($R_e$). This behavior has typically been observed for heteronuclear molecular ions and indicates that, unlike neutral molecules, molecular ions possess finite PDMs even at distances close to the dissociation limit~\cite{Fedorov_2017,Thakur_2025}. The magnitudes of the computed PDMs of the AlkH$^+$ molecular ions at $R_e$, obtained using different levels of theory, are tabulated in Table~\ref{table-IV}. We have found that the PDM at $R_e$ is negative for LiH$^+$, NaH$^+$, and KH$^+$, whereas it is positive for the remaining molecular ions and increases steadily from lighter to heavier systems. For LiH$^+$ and NaH$^+$, our recommended results at the CCSD(T) level are \(0.03\) atomic units (au) smaller than those reported in Ref.~\cite{Cheng_2007}, which employed a similar level of theory but smaller basis sets compared to ours. Relative to Ref.~\cite{Laszlo_1982}, our calculated PDM values show increments of \(0.017\) au and \(0.006\) au for RbH$^+$ and CsH$^+$, respectively.\\

The variations of the parallel ($\alpha_\parallel$) and perpendicular ($\alpha_\perp$) components of the static electric dipole polarizability with $R$ for the AlkH$^+$ ions are presented in Fig.~\ref{fig:FIG3} and Fig.~\ref{fig:FIG4}, respectively. $\alpha_\parallel$ decreases initially and then increases up to $R$\,=\,$R_e$ for LiH$^+$ through RbH$^+$, whereas it continues to decrease for CsH$^+$ and FrH$^+$. In contrast, $\alpha_\perp$ exhibits an increasing trend over the same range of $R$ from LiH$^+$ to RbH$^+$, except for NaH$^+$. Beyond $R_e$, both $\alpha_\parallel$ and $\alpha_\perp$ gently flatten out and approach the atomic dipole polarizability limit at large internuclear separations for all molecular ions. Further, the dipole polarizability components ($\alpha_\parallel, \alpha_\perp$, $\bar{\alpha}$ and $\gamma$) at $R_e$ are tabulated in Table~\ref{table-IV} and they increase from LiH$^+$ to FrH$^+$. The contributions of electron correlation effects beyond the DHF level to ($\bar{\alpha}$, $\gamma$) are (\(0.2\)\%, \(3.5\)\%) for LiH$^+$, (\(3\)\%, \(10\)\%) for NaH$^+$, (\(1.5\)\%, \(6\)\%) for KH$^+$, (\(1\)\%, \(13\)\%) for RbH$^+$, (\(0.01\)\%, \(29\)\%) for CsH$^+$, and (\(1.1\)\%, \(62\)\%) for FrH$^+$. Barring the lightest two members of the AlkH$^+$ series, the polarizabilities of the other systems have been reported for the first time in this work.\\

Table~\ref{table-V} lists the molecular polarizabilities ($\alpha_\parallel$ and $\alpha_\perp$) calculated at $R$\,=\,\(30\) {\AA} and the asymptotic values of the polarizabilities obtained using Eqs.~(\ref{eq:1}) and (\ref{eq:2}). We have found excellent agreement between the two sets of values, with deviations below \(0.8\)\% for all the molecular ions. This observation further confirms the identities of the dissociation products of the AlkH$^+$ molecular ions.\\

The QM values for all molecular ions increase progressively with increasing $R$ and attain approximately constant values beyond $R$\,$\approx$\,\(4\) {\AA}. This behavior is illustrated in Fig.~\ref{fig:FIG5}. The calculated QM values at $R_e$ are reported in Table~\ref{table-IV}. A systematic increase in QM is observed along the series from LiH$^+$ to FrH$^+$. Further, the largest electron correlation effects are found for the CsH$^+$ and FrH$^+$ ions, with contributions of \(0.6\)\% and \(0.5\)\%, respectively. To the best of our knowledge, no prior studies have reported QM results for the AlkH$^+$ molecular ions in the literature. Both PDMs and QMs play an important role in describing higher-order electrostatic interactions, which exhibit complex orientation dependencies involving dipole-dipole, dipole-quadrupole, and quadrupole-quadrupole terms~\cite{Walraven_2024}.\\

As we have already employed a large cv5z basis set, the basis set superposition error in the energy calculations, and consequently in the derived molecular properties, is expected to be minimal~\cite{Belpassi_2008,Miliordos_2015}. We have analyzed the effect of enlarging the correlation space for RbH$^+$ from the range of \(-5\) $E_h$ to \(13\) $E_h$ used in this work to an extended range of \(-13\) $E_h$ to \(28\) $E_h$. This expansion of the configurational space has yielded an additional \(8\) active electrons and \(29\) virtual orbitals, albeit at a significant computational cost. The resulting changes in the calculated $D_e$, PDM, and $\alpha_\parallel$ are \(2\)\%, \(0.2\)\%, and \(0.2\)\% respectively, at the CCSD(T) level of theory. Therefore, the computed spectroscopic constants and molecular properties for the other molecular ions are also expected to show similar sensitivity to the size of the configurational space.

\subsection{Vibrational parameters}

We have calculated all vibrational parameters using the PEC and PDM curves obtained at the CCSD(T) level with the cv5z basis set. The maximum number of bound vibrational levels is found to be \(5\) for LiH$^+$, \(4\) for NaH$^+$, and \(3\) for the rest of the molecular ions. This is a consequence of the shallow potential well depths of all these ions. Table~\ref{table-VI} presents the vibrational energies ($E_v$), vibrationally coupled rotational constants ($B_v$), spontaneous and BBR-induced transition rates, and lifetimes for all vibrational states of AlkH$^+$ molecular ions. The relative energy spacings obtained in this work for LiH$^+$ are in good agreement with those reported in Refs.~\cite{Grabandt_1992} and \cite{Berriche_1995}, with maximum differences of \(20\) cm$^{-1}$ and \(4\) cm$^{-1}$, respectively. Furthermore, the vibrational energy differences and TDMs between different vibrational levels have been utilized to compute the spontaneous and BBR-induced transition rates at room temperature ($\mathrm{T}$ = \(300\) $\mathrm{K}$). All calculated TDMs between vibrational states are provided in Supplementary Table S2. The vibrational state lifetimes have then been determined as the inverse of the total transition rates. As shown in Table~\ref{table-VI}, the lifetime of the given vibrational state increases as one progresses from LiH$^+$ to FrH$^+$ molecular ions. The intermediate vibrational state has the shortest lifetime for all AlkH$^+$ ions because both emission and absorption channels are open for this state.

\section{\label{sec:section4}Conclusion}

In summary, we have performed a comprehensive study of the spectroscopic constants and molecular properties of the ground electronic state of alkali metal hydride cations. For this purpose, the higher-order electron correlation CCSD(T) method has been employed within a four-component relativistic formalism along with a large quintuple-zeta basis set. We have also analyzed the behavior of molecular properties over a range of internuclear distances. Further, QMs for all the molecular ions have been reported here for the first time, to the best of our knowledge. We confirm that the AlkH$^+$ molecular ions dissociate into Alk$^+$ and H atoms at the dissociation limit from the excellent agreement between the molecular energies (molecular polarizabilities) at the dissociation limit and the sum of atomic energies (atomic polarizabilities). We have also examined the effect of enlarging the configurational space on the computed results, which turns out to be around \(2\)\% for RbH$^+$. Finally, we have calculated the vibrational wavefunctions, $E_v$, $B_v$, and spontaneous and BBR-induced transition rates. The corresponding lifetimes of the vibrational levels have also been determined and are found to increase from LiH$^+$ to FrH$^+$ molecular ions.\\

We believe that the accurate \textit{ab initio} results presented here may be of interest to experimentalists investigating these molecular systems and may also serve as benchmarks for precise calculations in the future.

\begin{acknowledgments}

We would like to acknowledge Kenneth G. Dyall for providing the relativistic quintuple-zeta basis sets prior to their online availability. We would like to thank the National Supercomputing Mission (NSM) for providing computing resources of ‘PARAM Ganga’ at the Indian Institute of Technology Roorkee, implemented by C-DAC and supported by the Ministry of Electronics and Information Technology (MeitY) and Department of Science and Technology (DST), Government of India. Some of the calculations reported in this work were performed on the high-performance computing facility available in the Department of Physics, IIT Roorkee, India. R.B. was supported by Polish National Science Centre Project No. 2021/41/B/ST2/00681. The research is also a part of the program of the National Laboratory FAMO in Toru\'n, Poland.

\end{acknowledgments}
\bibliography{AlKH}

@PREAMBLE{
 "\providecommand{\noopsort}[1]{}" 
 # "\providecommand{\singleletter}[1]{#1}%" 
}

@article{Reitsma_2025,
    author = {Reitsma, Marten L. and Prinsen, Eifion H. and Polet, Johan D. and Borschevsky, Anastasia and Dyall, Kenneth G.},
    title = {Relativistic quintuple-zeta basis sets for the s block},
    journal = {The Journal of Chemical Physics},
    volume = {163},
    number = {14},
    pages = {144116},
    year = {2025},
    month = {10},
    issn = {0021-9606},
    doi = {10.1063/5.0297009},
    url = {https://doi.org/10.1063/5.0297009}
}

@article{Yan_2012,
    author = {Yan, Lingling and Qu, Yizhi and Liu, Chunhua and Wang, Jianguo and Buenker, Robert J.},
    title = {\textit{Ab initio} many-electron study for the low-lying states of the alkali hydride cations in the adiabatic representation},
    journal = {The Journal of Chemical Physics},
    volume = {136},
    number = {12},
    pages = {124304},
    year = {2012},
    month = {03},
    issn = {0021-9606},
    doi = {10.1063/1.3695997},
    url = {https://doi.org/10.1063/1.3695997}
}

@article{Aymar_2009,
author = {Aymar, Mireille and Deiglmayr, Johannes and Dulieu, Olivier},
title = {Systematic trends in electronic properties of alkali hydrides},
journal = {Canadian Journal of Physics},
volume = {87},
number = {5},
pages = {543-556},
year = {2009},
doi = {10.1139/P09-005},
URL = {https://doi.org/10.1139/P09-005}
}

@article{Magnier_2004,
author = {Magnier, Sylvie},
title = {Electronic Structure of {LiH}$^+$ Revisited by a Model Potential-Type Method},
journal = {The Journal of Physical Chemistry A},
volume = {108},
number = {6},
pages = {1052-1056},
year = {2004},
doi = {10.1021/jp030833m},
URL = {https://doi.org/10.1021/jp030833m}
}

@article{Valance_1978,
title = {Pseudopotential calculations for {NaH}$^+$, {KH}$^+$, {RbH}$^+$ and {CsH}$^+$},
journal = {Chemical Physics Letters},
volume = {56},
number = {2},
pages = {289-294},
year = {1978},
issn = {0009-2614},
doi = {https://doi.org/10.1016/0009-2614(78)80242-5},
url = {https://www.sciencedirect.com/science/article/pii/0009261478802425},
author = {A. Valance}
}

@article{Watanabe_2002,
  title = {Charge-transfer cross sections for radiative charge transfer in {Na}+{H}$^{+}$ and {K}+{H}$^{+}$ collisions at very low energies},
  author = {Watanabe, Ayako and Dutta, C. M. and Nordlander, P. and Kimura, M. and Dalgarno, A.},
  journal = {Phys. Rev. A},
  volume = {66},
  issue = {4},
  pages = {044701},
  numpages = {4},
  year = {2002},
  month = {Oct},
  publisher = {American Physical Society},
  doi = {10.1103/PhysRevA.66.044701},
  url = {https://link.aps.org/doi/10.1103/PhysRevA.66.044701}
}

@article{Magnier_2005,
author = {Magnier, Sylvie},
title = {Theoretical Description of the Electronic Structure of the Alkali Hydride Cation {NaH}$^+$},
journal = {The Journal of Physical Chemistry A},
volume = {109},
number = {24},
pages = {5411-5414},
year = {2005},
doi = {10.1021/jp040711+},
URL = {https://doi.org/10.1021/jp040711+}
}

@article{Magnier_2006,
title = {Theoretical determination of the electronic structure of {KH}$^+$},
journal = {Chemical Physics},
volume = {326},
number = {2},
pages = {375-380},
year = {2006},
issn = {0301-0104},
doi = {https://doi.org/10.1016/j.chemphys.2006.02.015},
url = {https://www.sciencedirect.com/science/article/pii/S0301010406001261},
author = {Sylvie Magnier}
}

@article{Korek_2008,
    author = {Korek, M. and Hammoud, S. and Allouche, A. R. and Harb, T.},
    title = {Theoretical calculation of the low lying electronic states of the molecular ion {RbH}$^+$ with spin-orbit effects},
    journal = {The Journal of Chemical Physics},
    volume = {129},
    number = {20},
    pages = {204304},
    year = {2008},
    month = {11},
    issn = {0021-9606},
    doi = {10.1063/1.2992074},
    url = {https://doi.org/10.1063/1.2992074}
}

@article{Laszlo_1982,
title = {Pseudopotential calculations on {Rb}${^+_2}$, {Cs}${^+_2}$, {RbH}$^+$, {CsH}$^+$ and the mixed alkali dimer ions},
journal = {Chemical Physics Letters},
volume = {93},
number = {6},
pages = {555-559},
year = {1982},
issn = {0009-2614},
doi = {https://doi.org/10.1016/0009-2614(82)83728-7},
url = {https://www.sciencedirect.com/science/article/pii/0009261482837287},
author = {László {von Szentpály} and Patricio Fuentealba and Heinzwerner Preuss and Hermann Stoll}
}

@article{Rosmus_1977,
author = {Rosmus, P. and Meyer, W.},
title = {{PNO}–{CI} and {CEPA} studies of electron correlation effects. {IV}. {I}onization energies of the first and second row diatomic hydrides and the spectroscopic constants of their ions},
journal = {The Journal of Chemical Physics},
volume = {66},
number = {1},
pages = {13-19},
year = {1977},
month = {01},
doi = {10.1063/1.433661},
url = {https://doi.org/10.1063/1.433661}
}

@article{DIRAC,
      Title = {\textit {{DIRAC}, a relativistic ab initio electronic structure program, Release {DIRAC}23}},
      year= {2023 (http://www.diracprogram.org)},
      author= {R. Bast and A. S. P. Gomes and T. Saue and L. Visscher and H. J. Aa. Jensen and with contributions from I. A. Aucar and V. Bakken and C. Chibueze and J. Creutzberg and K. G. Dyall and S. Dubillard and U. Ekström and E. Eliav and T. Enevoldsen and E. Faßhauer and T. Fleig and O. Fossgaard and L. Halbert and E. D. Hedegård and T. Helgaker and B. Helmich-Paris and J. Henriksson and M. van Horn and M. Iliaš and Ch. R. Jacob and S. Knecht and S. Komorovský and O. Kullie and J. K. Lærdahl and C. V. Larsen and Y. S. Lee and N. H. List and H. S. Nataraj and M. K. Nayak and P. Norman and A. Nyvang and G. Olejniczak and J. Olsen and J. M. H. Olsen and A. Papadopoulos and Y. C. Park and J. K. Pedersen and M. Pernpointner and J. V. Pototschnig and R. Di Remigio and M. Repisky and K. Ruud and P. Sałek and B. Schimmelpfennig and B. Senjean and A. Shee and J. Sikkema and A. Sunaga and A. J. Thorvaldsen and J. Thyssen and J. van Stralen and M. L. Vidal and S. Villaume and O. Visser and T. Winther and S. Yamamoto and X. Yuan},
      journal= {},
     volume = {},
      pages = {},
      url= {http://www.diracprogram.org},
      note={available at \url{https://doi.org/10.5281/zenodo.7670749}}
}

@article{OPENMOLCAS,
title = {"{OpenMolcas: From Source Code to Insight}"},
author = {I. F. Galv\'an
and Morgane Vacher and Ali Alavi and Celestino Angeli and Francesco Aquilante and Jochen Autschbach and Jie J. Bao and Sergey I. Bokarev and Nikolay A. Bogdanov and Rebecca K. Carlson and Liviu F. Chibotaru and Joel Creutzberg and Nike Dattani and Mickaël G. Delcey and Sijia S. Dong and Andreas Dreuw and Leon Freitag and Luis Manuel Frutos and Laura Gagliardi and Frédéric Gendron and Angelo Giussani and Leticia González and Gilbert Grell and Meiyuan Guo and Chad E. Hoyer and Marcus Johansson and Sebastian Keller and Stefan Knecht and Goran Kovačević and Erik Källman and Giovanni Li Manni and Marcus Lundberg and Yingjin Ma and Sebastian Mai and João Pedro Malhado and Per Åke Malmqvist and Philipp Marquetand and Stefanie A. Mewes and Jesper Norell and Massimo Olivucci and Markus Oppel and Quan Manh Phung and Kristin Pierloot and Felix Plasser and Markus Reiher and Andrew M. Sand and Igor Schapiro and Prachi Sharma and Christopher J. Stein and Lasse Kragh Sørensen and Donald G. Truhlar and Mihkel Ugandi and Liviu Ungur and Alessio Valentini and Steven Vancoillie and Valera Veryazov and Oskar Weser and Tomasz A. Wesołowski and Per-Olof Widmark and Sebastian Wouters and Alexander Zech and J. Patrick Zobel and Roland Lindh.},
journal = {Journal of Chemical Theory and Computation},
volume = {15},
number = {11},
pages = {5925-5964},
year = {2019},
doi = {10.1021/acs.jctc.9b00532},
URL = {https://doi.org/10.1021/acs.jctc.9b00532}
}

@article{LEVEL,
title = {{LEVEL}: A computer program for solving the radial {S}chr\"odinger equation for bound and quasibound levels},
journal = {Journal of Quantitative Spectroscopy and Radiative Transfer},
volume = {186},
pages = {167-178},
year = {2017},
issn = {0022-4073},
doi = {https://doi.org/10.1016/j.jqsrt.2016.05.028},
url = {https://www.sciencedirect.com/science/article/pii/S0022407316300978},
author = {Robert J. {Le Roy}}
}

@article{Maroulis_1998,
    author = {Maroulis, George},
    title = {A systematic study of basis set, electron correlation, and geometry effects on the electric multipole moments, polarizability, and hyperpolarizability of {HCl}},
    journal = {The Journal of Chemical Physics},
    volume = {108},
    number = {13},
    pages = {5432-5448},
    year = {1998},
    month = {04},
    issn = {0021-9606},
    doi = {10.1063/1.475932},
    url = {https://doi.org/10.1063/1.475932}
}

@article{Thakur_2024,
doi = {10.1088/1361-6455/ad7e8b},
url = {https://dx.doi.org/10.1088/1361-6455/ad7e8b},
year = {2024},
month = {oct},
publisher = {IOP Publishing},
volume = {57},
number = {21},
pages = {215102},
author = {Thakur, Ankush and Bala, Renu and Nataraj, H S and Prasannaa, V S},
title = {Relativistic coupled-cluster calculations for the molecular properties of {AlX}$^+$ ({X}: {F}, {Cl}, {Br}, {I}, {At} and {Ts}) ions},
journal = {Journal of Physics B: Atomic, Molecular and Optical Physics}
}

@Article{Buckingham_1959,
author ="Buckingham, A. D.",
title  ="Molecular quadrupole moments",
journal  ="Q. Rev. Chem. Soc.",
year  ="1959",
volume  ="13",
issue  ="3",
pages  ="183-214",
publisher  ="The Royal Society of Chemistry",
doi  ="10.1039/QR9591300183",
url  ="http://dx.doi.org/10.1039/QR9591300183"}

@article{Jensen_2002,
    author = {Jensen, Lasse and Åstrand, Per-Olof and Osted, Anders and Kongsted, Jacob and Mikkelsen, Kurt V.},
    title = {Polarizability of molecular clusters as calculated by a dipole interaction model},
    journal = {The Journal of Chemical Physics},
    volume = {116},
    number = {10},
    pages = {4001-4010},
    year = {2002},
    month = {03},
    issn = {0021-9606},
    doi = {10.1063/1.1433747},
    url = {https://doi.org/10.1063/1.1433747}
}

@article{Ghanmi_2025, 
title={Investigation of the {FrH}$^+$ Alkali Hydride Cation: Analysis of Potential Energies, Dipole Functions, and Radiative Lifetimes},
journal={Turkish Computational and Theoretical Chemistry},
volume={9},
pages={79–88},
year={2025},
DOI={10.33435/tcandtc.1418065},
url={https://izlik.org/JA76SG67PZ},
author={Ghanmi, Chedli and Berriche, Hamid},
number={3}
}

@article{Cheng_2007,
  title = {Dipole moments and orientation polarizabilities of diatomic molecular ions for precision atomic mass measurement},
  author = {Cheng, Michelle and Brown, John M. and Rosmus, Pavel and Linguerri, Roberto and Komiha, Najia and Myers, Edmund G.},
  journal = {Phys. Rev. A},
  volume = {75},
  issue = {1},
  pages = {012502},
  numpages = {14},
  year = {2007},
  month = {Jan},
  publisher = {American Physical Society},
  doi = {10.1103/PhysRevA.75.012502},
  url = {https://link.aps.org/doi/10.1103/PhysRevA.75.012502}
}

@article{Fedorov_2017,
    author = {Fedorov, Dmitry A. and Barnes, Dustin K. and Varganov, Sergey A.},
    title = {\textit{Ab initio} calculations of spectroscopic constants and vibrational state lifetimes of diatomic alkali-alkaline-earth cations},
    journal = {The Journal of Chemical Physics},
    volume = {147},
    number = {12},
    pages = {124304},
    year = {2017},
    month = {09},
    issn = {0021-9606},
    doi = {10.1063/1.4986818},
    url = {https://doi.org/10.1063/1.4986818}
}

@article{Thakur_2025,
title = {\textit{Ab initio} calculations of diatomic constants and ro-vibrational parameters for the ground state of singly charged aluminium monohalides},
journal = {Journal of Quantitative Spectroscopy and Radiative Transfer},
volume = {345},
pages = {109574},
year = {2025},
issn = {0022-4073},
doi = {https://doi.org/10.1016/j.jqsrt.2025.109574},
url = {https://www.sciencedirect.com/science/article/pii/S0022407325002365},
author = {Ankush Thakur and Renu Bala and H. S. Nataraj}
}

@article{Grabandt_1992,
title = {The temperature dependence of the {LiH} photoelectron spectrum},
journal = {Chemical Physics Letters},
volume = {189},
number = {4},
pages = {291-296},
year = {1992},
issn = {0009-2614},
doi = {https://doi.org/10.1016/0009-2614(92)85204-N},
url = {https://www.sciencedirect.com/science/article/pii/000926149285204N},
author = {O. Grabandt and H.J. Bakker and C.A. {de Lange}}
}

@article{Berriche_1995,
title = {Ab initio study of the {LiH}$^+$ molecule, electronic interaction analysis and {LiH} {UV} photoelectron spectrum},
journal = {Chemical Physics},
volume = {191},
number = {1},
pages = {119-131},
year = {1995},
issn = {0301-0104},
doi = {https://doi.org/10.1016/0301-0104(94)00342-8},
url = {https://www.sciencedirect.com/science/article/pii/0301010494003428},
author = {Hamid Berriche and Florent Xavier Gadea}
}

@article{Kotochigova_2005,
    author = {Kotochigova, S. and Tiesinga, E.},
    title = "{\textit{Ab initio} relativistic calculation of the {RbCs} molecule}",
    journal = {The Journal of Chemical Physics},
    volume = {123},
    number = {17},
    pages = {174304},
    year = {2005},
    month = {10},
    issn = {0021-9606},
    doi = {10.1063/1.2107607},
    url = {https://doi.org/10.1063/1.2107607}
}

@book{Levine_1985,
  author       = {Levine, J S},
  title        = {Photochemistry of atmospheres: Earth, the other planets, and comets},
  url          = {https://www.osti.gov/biblio/5895573},
  place        = {United States},
  publisher    = {Academic Press, Orlando, FL},
  year         = {1985},
  month        = {01}}

@article{Langer_1990,
  author  = {Langer, G. E. and Prosser, Charles F. and Sneden, Christopher},
  title   = {Variations in Cold Interstellar Clouds on Subparsec Scales},
  journal = {The Astronomical Journal},
  volume  = {100},
  number  = {1},
  pages   = {216--222},
  year    = {1990},
  doi  =  {10.1086/115506}
}

@article{Watson_1996,
doi = {10.1086/310410},
url = {https://doi.org/10.1086/310410},
year = {1996},
month = {dec},
publisher = {},
volume = {473},
number = {2},
pages = {L127},
author = {Watson, John K. and Meyer, David M.},
title = {Observations of Ubiquitous Small-Scale Structure in the Diffuse Interstellar Medium},
journal = {The Astrophysical Journal}
}

@article{Burrows_2001,
  title = {The theory of brown dwarfs and extrasolar giant planets},
  author = {Burrows, Adam and Hubbard, W. B. and Lunine, J. I. and Liebert, James},
  journal = {Rev. Mod. Phys.},
  volume = {73},
  issue = {3},
  pages = {719--765},
  numpages = {0},
  year = {2001},
  month = {Sep},
  publisher = {American Physical Society},
  doi = {10.1103/RevModPhys.73.719},
  url = {https://link.aps.org/doi/10.1103/RevModPhys.73.719}
}

@article{Krohn_2001,
  title = {Rate Coefficients and Final States for the Dissociative Recombination of $\mathrm{LiH}^{+}$},
  author = {Krohn, S. and Lange, M. and Grieser, M. and Knoll, L. and Kreckel, H. and Levin, J. and Repnow, R. and Schwalm, D. and Wester, R. and Witte, P. and Wolf, A. and Zajfman, D.},
  journal = {Phys. Rev. Lett.},
  volume = {86},
  issue = {18},
  pages = {4005--4008},
  numpages = {0},
  year = {2001},
  month = {Apr},
  publisher = {American Physical Society},
  doi = {10.1103/PhysRevLett.86.4005},
  url = {https://link.aps.org/doi/10.1103/PhysRevLett.86.4005}
}

@article{Hack_1999,
    author = {Hack, Michael D. and Truhlar, Donald G.},
    title = {Analytic potential energy surfaces and their couplings for the electronically nonadiabatic chemical processes {Na(3p)} + {H$_2$} $\rightarrow$ {Na(3s)} + {H$_2$} and {Na(3p)} + {H$_2$} $\rightarrow$ {NaH} + {H}},
    journal = {The Journal of Chemical Physics},
    volume = {110},
    number = {9},
    pages = {4315-4337},
    year = {1999},
    month = {03},
    issn = {0021-9606},
    doi = {10.1063/1.478314},
    url = {https://doi.org/10.1063/1.478314}
}

@article{Lin_1989,
    author = {Lin, King‐Chuen and Chang, Hai‐Chou},
    title = {State‐selective reaction of excited potassium atom with hydrogen molecule. {K}$^{*}$ + {H$_2$} $\rightarrow$ {KH} + {H}},
    journal = {The Journal of Chemical Physics},
    volume = {90},
    number = {11},
    pages = {6151-6156},
    year = {1989},
    month = {06},
    issn = {0021-9606},
    doi = {10.1063/1.456331},
    url = {https://doi.org/10.1063/1.456331}
}

@article{Wang_2025,
  title = {Photodissociation dynamics of ${X}${H}$^{+}$ (${X}$ = {Na}, {K}, {Rb}, {Cs}): Rovibrationally resolved shape resonances driven by potential barriers in the ultraviolet spectrum},
  author = {Wang, Kun and Su, Xing Guang and Liu, Dong and Wang, Yu and Cheng, Yongjun and Wu, Yong and Wang, Jian Guo and Qu, Yi Zhi and Wang, Ke Dong and Zhang, Song Bin and Yang, Yu Kun},
  journal = {Phys. Rev. A},
  volume = {112},
  issue = {1},
  pages = {013119},
  numpages = {9},
  year = {2025},
  month = {Jul},
  publisher = {American Physical Society},
  doi = {10.1103/ny9t-wqp1},
  url = {https://link.aps.org/doi/10.1103/ny9t-wqp1}
}

@article{Cote_2000,
  title = {Enhanced Cooling of Hydrogen Atoms by Lithium Atoms},
  author = {C\^ot\'e, R. and Jamieson, M. J. and Yan, Z-C. and Geum, N. and Jeung, G.-H. and Dalgarno, A.},
  journal = {Phys. Rev. Lett.},
  volume = {84},
  issue = {13},
  pages = {2806--2809},
  numpages = {0},
  year = {2000},
  month = {Mar},
  publisher = {American Physical Society},
  doi = {10.1103/PhysRevLett.84.2806},
  url = {https://link.aps.org/doi/10.1103/PhysRevLett.84.2806}
}

@article{Derevianko_2001,
  title = {Enhanced cooling of hydrogen by a buffer gas of alkali-metal atoms},
  author = {Derevianko, A. and C\^ot\'e, R. and Dalgarno, A. and Jeung, G.-H.},
  journal = {Phys. Rev. A},
  volume = {64},
  issue = {1},
  pages = {011404(R)},
  numpages = {4},
  year = {2001},
  month = {Jun},
  publisher = {American Physical Society},
  doi = {10.1103/PhysRevA.64.011404},
  url = {https://link.aps.org/doi/10.1103/PhysRevA.64.011404}
}

@article{Scheidt_1978,
doi = {10.1088/0022-3700/11/15/015},
url = {https://doi.org/10.1088/0022-3700/11/15/015},
year = {1978},
month = {aug},
publisher = {},
volume = {11},
number = {15},
pages = {2665},
author = {H Scheidt and G Spiess and A Valance and P Pradel},
title = {Determination of {H}$^+$+\,{Cs}(6s) potential from differential cross section measurements at energies 13.4-24.2 {eV}},
journal = {Journal of Physics B: Atomic and Molecular Physics}
}

@article{Thakur_2026,
author    = {Ankush Thakur and Renu Bala and H. S. Nataraj},
title     = {Study of ground state electronic structure of {XH}$^{+}$ ({X}: {Cd}, {Hg}, and {Yb}) molecular ions via coupled-cluster approach},
journal   = {The European Physical Journal Plus},
year      = {2026},
volume    = {141},
number    = {4},
pages     = {399},
doi       = {10.1140/epjp/s13360-026-07630-8},
url       = {https://doi.org/10.1140/epjp/s13360-026-07630-8},
issn      = {2190-5444}
}

@article{Miliordos_2015,
    author = {Miliordos, Evangelos and Xantheas, Sotiris S.},
    title = {On the validity of the basis set superposition error and complete basis set limit extrapolations for the binding energy of the formic acid dimer},
    journal = {The Journal of Chemical Physics},
    volume = {142},
    number = {9},
    pages = {094311},
    year = {2015},
    month = {03},
    issn = {0021-9606},
    doi = {10.1063/1.4913766},
    url = {https://doi.org/10.1063/1.4913766}
}

@article{Belpassi_2008,
author = {Belpassi, Leonardo and Infante, Ivan and Tarantelli, Francesco and Visscher, Lucas},
title = {The {C}hemical {B}ond between {Au(I)} and the {N}oble {G}ases. {C}omparative {S}tudy of {NgAuF} and {NgAu}$^+$ ({Ng} = {Ar}, {Kr}, {Xe}) by {D}ensity {F}unctional and {C}oupled {C}luster {M}ethods},
journal = {Journal of the American Chemical Society},
volume = {130},
number = {3},
pages = {1048-1060},
year = {2008},
doi = {10.1021/ja0772647},
URL = {https://doi.org/10.1021/ja0772647}
}

@article{Silberstein_1917,
author = {L. Silberstein},
title = {L. {M}olecular refractivity and atomic interaction. {II}},
journal = {The London, Edinburgh, and Dublin Philosophical Magazine and Journal of Science},
volume = {33},
number = {198},
pages = {521--533},
year = {1917},
publisher = {Taylor \& Francis},
doi = {10.1080/14786440608635666},
URL = {https://doi.org/10.1080/14786440608635666}
}

@article{Sharp_2007,
doi = {10.1086/508708},
url = {https://doi.org/10.1086/508708},
year = {2007},
month = {jan},
publisher = {},
volume = {168},
number = {1},
pages = {140},
author = {Sharp, C. M. and Burrows, A.},
title = {Atomic and Molecular Opacities for Brown Dwarf and Giant Planet Atmospheres},
journal = {The Astrophysical Journal Supplement Series}
}

@article{Zajfman_2003,
  author  = {Zajfman, D. and Schwalm, D. and Wolf, A.},
  title   = {Molecular Physics in Storage Rings: From Laboratory to Space},
  journal = {Hyperfine Interactions},
  volume  = {146},
  number  = {1},
  pages   = {265--268},
  year    = {2003},
  month   = mar,
  doi     = {10.1023/B:HYPE.0000004250.68609.53},
  url     = {https://doi.org/10.1023/B:HYPE.0000004250.68609.53}
}

@article{Walraven_2024,
  title = {Rotational-state dependence of interactions between polar molecules},
  author = {Walraven, Etienne F. and Karman, Tijs},
  journal = {Phys. Rev. A},
  volume = {109},
  issue = {4},
  pages = {043310},
  numpages = {12},
  year = {2024},
  month = {Apr},
  publisher = {American Physical Society},
  doi = {10.1103/PhysRevA.109.043310},
  url = {https://link.aps.org/doi/10.1103/PhysRevA.109.043310}
}

\end{document}